\documentclass[amsmath,amssymb,twocolumn,prl,aps,floatfix]{revtex4-1}
\usepackage{comment}
\usepackage{graphicx,ams math,epsfig,epstopdf}
\usepackage{dcolumn}
\usepackage{bm}
\usepackage{hyperref}
\usepackage{tabularx}
\usepackage{color}
\usepackage{mathrsfs}
\usepackage{ulem}
\hypersetup{
	colorlinks = True,
	urlcolor=blue,
	linkcolor = blue,
	citecolor= blue
}
\usepackage{algorithm}
\usepackage{algpseudocode}

\begin{document}

\title{A Neural Decoder for Topological Codes}%

\author{Giacomo Torlai and Roger G. Melko}
\address{
Department of Physics and Astronomy, University of Waterloo, Ontario N2L 3G1, Canada\\
Perimeter Institute of Theoretical Physics, Waterloo, Ontario N2L 2Y5, Canada
}
\date{\today}

\begin{abstract}
We present an algorithm for error correction in topological codes that exploits modern machine learning techniques. Our decoder is constructed from a stochastic neural network called a Boltzmann machine, of the type extensively used in deep learning. We provide a general prescription for the training of the network and a decoding strategy that is applicable to a wide variety of stabilizer codes with very little specialization.  We demonstrate the neural decoder numerically on the well-known two dimensional toric code with phase-flip errors.

\end{abstract}
\maketitle

{\it Introduction}: 
Much of the success of modern machine learning stems from the flexibility of a given neural network architecture to be employed for a multitude of different tasks. 
This generalizability means that neural networks can have the ability to infer structure from vastly different data sets with only a change in optimal 
hyper-parameters.
For this purpose, the machine learning community has developed a set of standard tools, such as fully-connected feed forward networks~\cite{Hornik89} and Boltzmann machines~\cite{Salakhutdinov08_BM}.
Specializations of these underlie many of the more advanced algorithms, including convolutional networks~\cite{Krizhevsky12} and deep learning~\cite{Hinton07,LeCun15},
encountered in real-world applications such as image or speech recognition~\cite{Hinton12speech}.

These machine learning techniques may be harnessed for a multitude of complex tasks in science and engineering~\cite{Torlai,Wang16,Carrasquilla16,Broecker16,Ch'ng16,Carleo16,Deng16,Huang16,Liu16,Stoudenmire16}.  An important application lies in 
quantum computing.  For a quantum logic operation to succeed, noise sources which lead to decoherence in a qubit must be mitigated.
This can be done through some type of quantum error correction -- a process where the logical state of a qubit is encoded redundantly so that errors can be corrected before they corrupt it.
A leading candidate for this is the implementation of fault-tolerant hardware through {\it surface codes}, where a logical qubit is stored as 
a topological state of an array of {\it physical} qubits~\cite{Bombin13}.
Random errors in the states of the physical qubits can be corrected before they proliferate and
destroy the logical state.  The quantum error correction protocols that perform this correction are termed ``decoders'', and must be implemented by 
classical algorithms running on conventional computers~\cite{Fowler12PRA}.
 
In this paper we demonstrate how one of the simplest stochastic neural networks for unsupervised learning, the restricted Boltzmann machine~\cite{Hinton12}, can be used to construct a general error-correction protocol for stabilizer codes.
Give a {\it syndrome}, defined by a measurement of the end points of an (unknown) chain of physical qubit errors,  
we use our Boltzmann machine to devise a protocol with the goal of correcting errors without corrupting the logical bit.
Our decoder works for generic degenerate stabilizers codes that have a probabilistic relation between syndrome and errors, which does not have to be {\it a priori}  known.  Importantly, it is very simple to implement, requiring no specialization regarding code locality, dimension, or structure.
We test our decoder numerically on a simple two-dimensional surface code with phase-flip errors.

\begin{figure}[t]
\centering
\includegraphics[width=60mm]{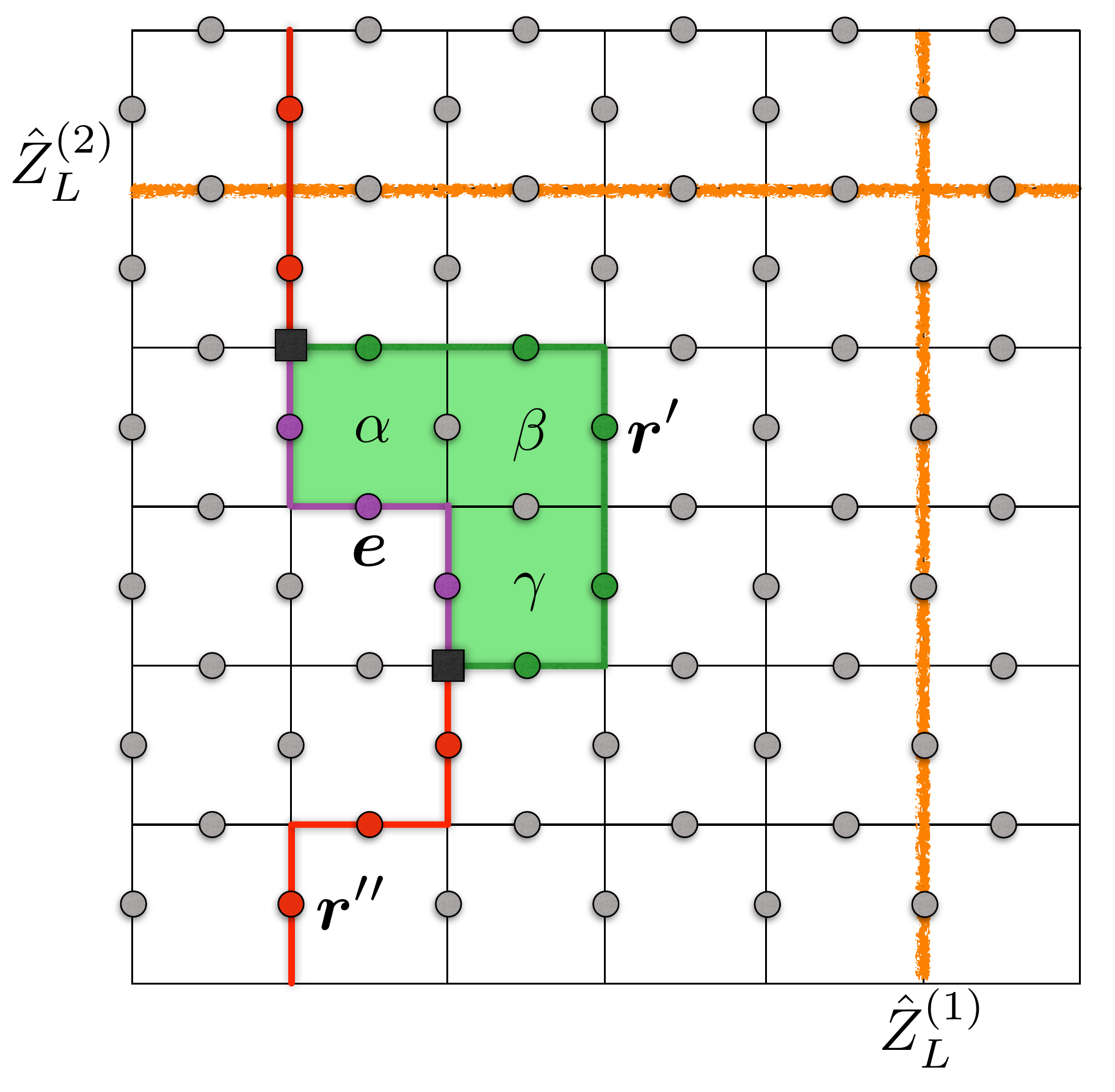}
\caption{Several operations on a $2D$ toric code. Logical operators $\hat{Z}_L^{(1)}$ and $\hat{Z}_L^{(1)}$ (orange) are non-trivial cycles on the real lattice. A physical error chain $\bm{e}$ (purple) and its syndrome $S(\bm{e})$ (black squares). A recovery chain $\bm{r}^\prime$ (green), with the combined operator on the cycle $\bm{e}\oplus\bm{r}^\prime$ being a product of stabilizers $\hat{Z}_\alpha\hat{Z}_\beta\hat{Z}_\gamma$ (recovery success). A recovery chain $\bm{r}^{\prime\prime}$ (red) whose cycle has non-trivial homology and acts on the code state as $\hat{Z}_L^{(1)}$ (logical failure). }
\label{TC}
\end{figure}

{\it The ${\rm 2D}$ Toric Code.} 
Most topological codes can be described in terms of the stabilizer formalism~\cite{Gottesman97}. 
A stabilizer code is a particular class of error-correcting code characterized by a protected subspace $\mathcal{C}$ defined by a stabilizer group $\mathcal{S}$.
The simplest example is the $2D$ toric code, first introduced by Kitaev~\cite{Kitaev03}.  Here, the quantum information is encoded into the homological degrees of freedom, with topological invariance given by the first homology group~\cite{Dennis02}. The code features $N$ qubits placed on the links of a $L\times L$ square lattice embedded on a torus. The stabilizers group is $\mathcal{S}=\{\hat{Z}_p,\hat{X}_v\}$, where the plaquette and vertex stabilizers are defined respectively as $\hat{Z}_p=\bigotimes_{\ell\in p}\,\hat{\sigma}^z_\ell$ and $\hat{X}_v=\bigotimes_{\ell\in v}\,\hat{\sigma}^x_\ell$, with $\hat{\sigma}^z_\ell$ and $\hat{\sigma}^x_\ell$ acting respectively on the links contained in the plaquette $p$ and the links connected to the vertex $v$. There are two encoded logical qubits, manipulated by logical operators $\hat{Z}_L^{(1,2)}$ as $\hat{\sigma}^z$ acting on the non-contractible loops on the real lattice and logical $\hat{X}_L^{(1,2)}$ as the non-contractible loops on the dual lattice (Fig~\ref{TC}).

Given a reference state $|\psi_0\rangle\in\mathcal{C}$, let us consider the simple phase-flip channel described by a Pauli operator 
where $\hat{\sigma}^z$ is applied to each qubit with probability $p_{err}$. This operator 
can be efficiently described by a mapping between the links and $\mathbb{Z}_2$, called an error chain $\bm{e}$, whose boundary is called a syndrome $\bm{S}(\bm{e})$. In a experimental implementation, only the syndrome (and not the error chain) can be measured.
Error correction (decoding) consists of applying a recovery operator 
whose chain $\bm{r}$ generates the same syndrome, $\bm{S}(\bm{e})=\bm{S}(\bm{r})$. The recovery succeeds only if the combined operation is described by a {\it cycle} (i.e.~a chain with no boundaries) $\bm{e}\oplus\bm{r}$ that belongs to the trivial homology class $h_0$, describing contractable loops on the torus. On the other hand, if the cycle belongs to a non-trivial homology class (being non-contractible on the torus), 
the recovery operation directly manipulates the encoded logical information, leading to a logical failure (Fig~\ref{TC}). 

Several decoders have been proposed for the $2D$ toric code, based on different strategies~\cite{Poulin10,Poulin14,Wootton12,Hutter14,Fowler13}. 
Maximum likelihood decoding consists of finding a recovery chain $\bm{r}$ with the most likely homology class~\cite{Bravyi14,Hastings16}. 
A different recovery strategy, designed to reduce computational complexity,  
consists of generating the recovery chain $\bm{r}$ compatible with the syndrome simply by using the minimum number of errors. 
Such a procedure, called Minimum Weight Perfect Matching~\cite{Edmonds65} (MWPM), has the advantage that can be performed without the knowledge of the error probability $p_{err}$. 
This algorithm is however sub-optimal (with lower threshold probability~\cite{Dennis02}) since it does not take into account the high degeneracy of the error chains given a syndrome.

\begin{figure}[t]
\centering
\includegraphics[width=70mm]{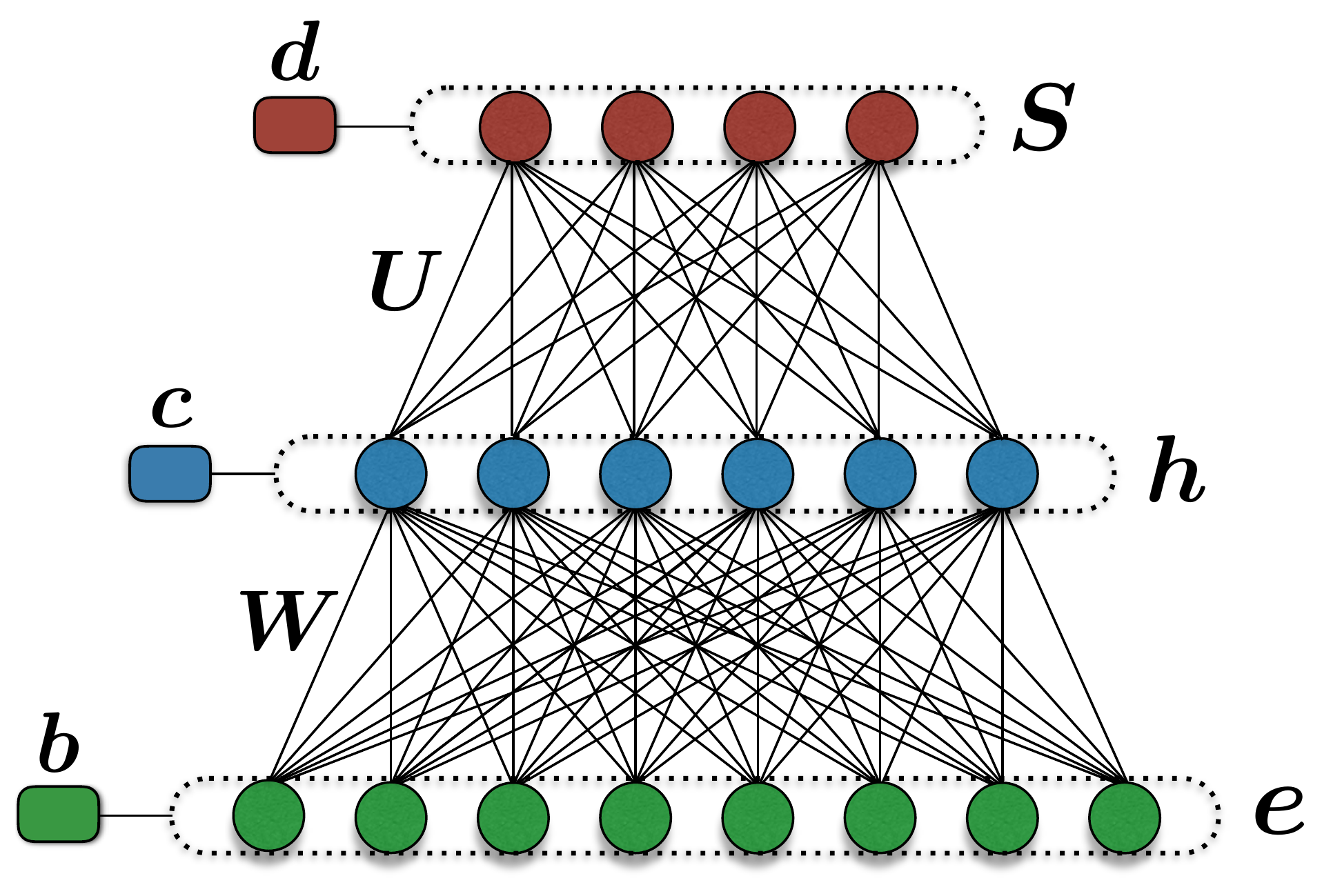}
\caption{The neural decoder architecture. The hidden layer $\bm{h}$ is fully-connected to the syndrome and error layers $\bm{S}$ and $\bm{e}$ with weights $\bm{U}$ and $\bm{W}$ respectively.}
\label{rbm}
\end{figure}

{\it The Neural Decoder.} Neural networks are commonly used to extract features from raw data in terms of probability distributions. In order to exploit this for error correction, we first build a dataset made of error chains and their syndromes $\mathcal{D}=\{\bm{e},\bm{S}\}$, and train a neural network to model the underlying probability distribution $p_{data}(\bm{e},\bm{S})$.  Our goal is to then generate error chains to use for the recovery.  We use a generative model called a Boltzmann machine, a powerful stochastic neural network widely used in the pre-training of the layers of deep neural networks~\cite{Hinton06,Salakhutdinov08}. The network architecture features three layers of stochastic binary neurons, the syndrome layer $\bm{S}\in\{0,1\}^{N/2}$, the error layer $\bm{e}\in\{0,1\}^{N}$, and one hidden layer $\bm{h}\in\{0,1\}^{n_h}$ (Fig.~\ref{rbm}). Symmetric edges connect both the syndrome and the error layer with the hidden layer.
We point out the this network is equivalent to a traditional bilayer restricted Boltzmann machine, where we have here divided the visible layer into two separate layers for clarity. The weights on the edges connecting the network layers are given by the matrices $\bm{U}$ and $\bm{W}$ with zero diagonal. Moreover, we also add external fields $\bm{b}$, $\bm{c}$ and $\bm{d}$ coupled to the every neuron in each layer. The probability distribution that the probabilistic model associates to this graph structure is the Boltzmann distribution~\cite{Fischer12}
\begin{equation}
p_{\bm{\lambda}}(\bm{e},\bm{S},\bm{h})=\frac{1}{Z_{\bm{\lambda}}}\mbox{e}^{-E_{\bm{\lambda}}(\bm{e},\bm{S},\bm{h})}
\end{equation}
where $Z_{\bm{\lambda}}=\mbox{Tr}_{\{\bm{h},\bm{S},\bm{E}\}}\,\,\mbox{e}^{-E_{\bm{\lambda}}(\bm{e},\bm{S},\bm{h})}$ is the partition function, $\bm{\lambda}=\{\bm{U},\bm{W},\bm{b},\bm{c},\bm{d}\}$ is the set of parameters of the model, and the energy is
\begin{equation}
\begin{split}
E_{\bm{\lambda}}(\bm{e},\bm{S},\bm{h}) &=-\sum_{ik}U_{ik}h_iS_k-\sum_{ij}W_{ij}h_ie_j+\\
&\quad\,-\sum_j b_j e_j-\sum_i c_i h_i-\sum_k d_k S_k .
\end{split}
\end{equation}
The joint probability distribution over $(\bm{e},\bm{S})$ is obtained after integrating out the hidden variables from the full distribution
\begin{equation}
p_{\bm{\lambda}}(\bm{e},\bm{S})=\sum_{\bm{h}}p_{\bm{\lambda}}(\bm{e},\bm{S},\bm{h})=\frac{1}{Z_{\bm{\lambda}}}\mbox{e}^{-\mathcal{E}_{\bm{\lambda}}(\bm{e},\bm{S})}
\end{equation}
where the effective energy $\mathcal{E}_{\bm{\lambda}}(\bm{e},\bm{S})$ can be computed exactly. Moreover, given the structure of the network, the conditional probabilities $p_{\bm{\lambda}}(\bm{e}\,|\,\bm{h})$, $p_{\bm{\lambda}}(\bm{S}\,|\,\bm{h})$ and $p_{\bm{\lambda}}(\bm{h}\,|\,\bm{e},\bm{S})$ are also known exactly. The training of the machine consists of tuning the parameters $\bm{\lambda}$ until the model probability $p_{\bm{\lambda}}(\bm{e},\bm{S})$ becomes close to the target distribution $p_{data}(\bm{E},\bm{S})$ of the dataset. This translates into solving an optimization problem over the parameters $\bm{\lambda}$ by minimizing the distance between the two distribution, defined as the Kullbach-Leibler (KL) divergence, $\mathbb{KL}\propto-\sum_{(\bm{e},\bm{S})\in\mathcal{D}}\,\log p_{\bm{\lambda}}(\bm{e},\bm{S})$. Details about the Boltzmann machine and its training algorithm are reported in the Supplementary Materials.

We now discuss the decoding algorithm, which proceeds assuming that we successfully learned the distribution $p_{\bm{\lambda}}(\bm{e},\bm{S})$. Given an error chain $\bm{e}_0$ with syndrome $\bm{S}_0$ we wish to use the Boltzmann machine to generate an error chain compatible with $\bm{S}_0$ to use for the recovery.
To achieve this goal we separately train networks on different datasets obtained from different error regimes $p_{err}$.
Assuming we know the error regimes that generated $\bm{e}_0$, the recovery procedure consists of sampling a recovery chain from the distribution $p_{\bm{\lambda}}(\bm{e}\,|\,\bm{S}_0)$ given by the network trained at the same probability $p_{err}$ of $\bm{e}_0$. 
Although the Boltzmann machine does not learn this distribution directly, by sampling the error and hidden layers while keeping the syndrome layer fixed to $\bm{S}_0$, since $p_{\bm{\lambda}}(\bm{e},\bm{S}_0)= p_{\bm{\lambda}}(\bm{e}\,|\,\bm{S}_0)p(\bm{S}_0)$, we are enforcing sampling from the desired conditional distribution.  An advantage of this procedure over decoders that employ conventional Monte Carlo \cite{Wootton12,Hutter14} on specific stabilizer codes is that specialized sampling algorithms tied to the stabilizer structure, or multi-canonical methods such as parallel tempering, are not required.

An error correction procedure can be defined as follows (Alg.~\ref{Decoder}): we first initialize the machine into a random state of the error and hidden layers (see Fig.~\ref{rbm}) and to $\bm{S}_0$ for the syndrome layer. We then let the machine equilibrate by repeatedly performing block Gibbs sampling. After a some amount of equilibration steps, we begin checking the syndrome of the error state $\bm{e}$ in the machine and, as soon as $\bm{S}(\bm{e})=\bm{S}_0$ we select it for the recovery operation.

\begin{algorithm}[H]
\caption{Neural Decoding Strategy}
\label{Decoder}
\begin{algorithmic}[1]
\State $\bm{e}_0$: physical error chain 
\State $\bm{S}_0=\bm{S}(\bm{e}_0)$\Comment{Syndrome Extraction}
\State RBM = $\{\bm{e},\bm{S}=\bm{S}_0,\bm{h}\}$\Comment{Network Initialization}
\While{$\bm{S}(\bm{e})\ne\bm{S}_0$}\Comment{Sampling}
\State Sample $\bm{h}\sim p(\bm{h}\,|\,\bm{e},\bm{S}_0)$
\State Sample $\bm{e}\sim p(\bm{e}\,|\,\bm{h})$
\EndWhile
\State $\bm{r}=\bm{e}$\Comment{Decoding}
\end{algorithmic}
\end{algorithm}
{\it Results.} 
We train neural networks in different error regimes by building several datasets $\mathcal{D}_p=\{\bm{e}_k,\bm{S}_k\}_{k=1}^M$ at elementary error probabilities $p=\{0.5,0.6,\dots,0.15\}$ of the phase-flip channel.
For a given error probability, the network hyper-parameters are individually optimized via a grid search (for details see the Supplementary Material).
Once training is complete, we perform decoding following the procedure laid out in Alg.~\ref{Decoder}. We generate a test set  $\mathcal{T}_p=\{\bm{e}_k\}_{k=1}^{M}$ and 
for each error chain $\bm{e}_k\in\mathcal{T}_p$, after a suitable equilibration time (usually $N_{eq}\propto10^2$ sampling steps), 
we collect the first error chain $\bm{e}$ compatible with the original syndrome, $\bm{S}(\bm{e}) = \bm{S}(\bm{e}_k)$.  
We use this error chain for the recovery, $\bm{r}^{(k)}=\bm{e}$. 
\begin{figure}[t]
\centering
\includegraphics[width=80mm]{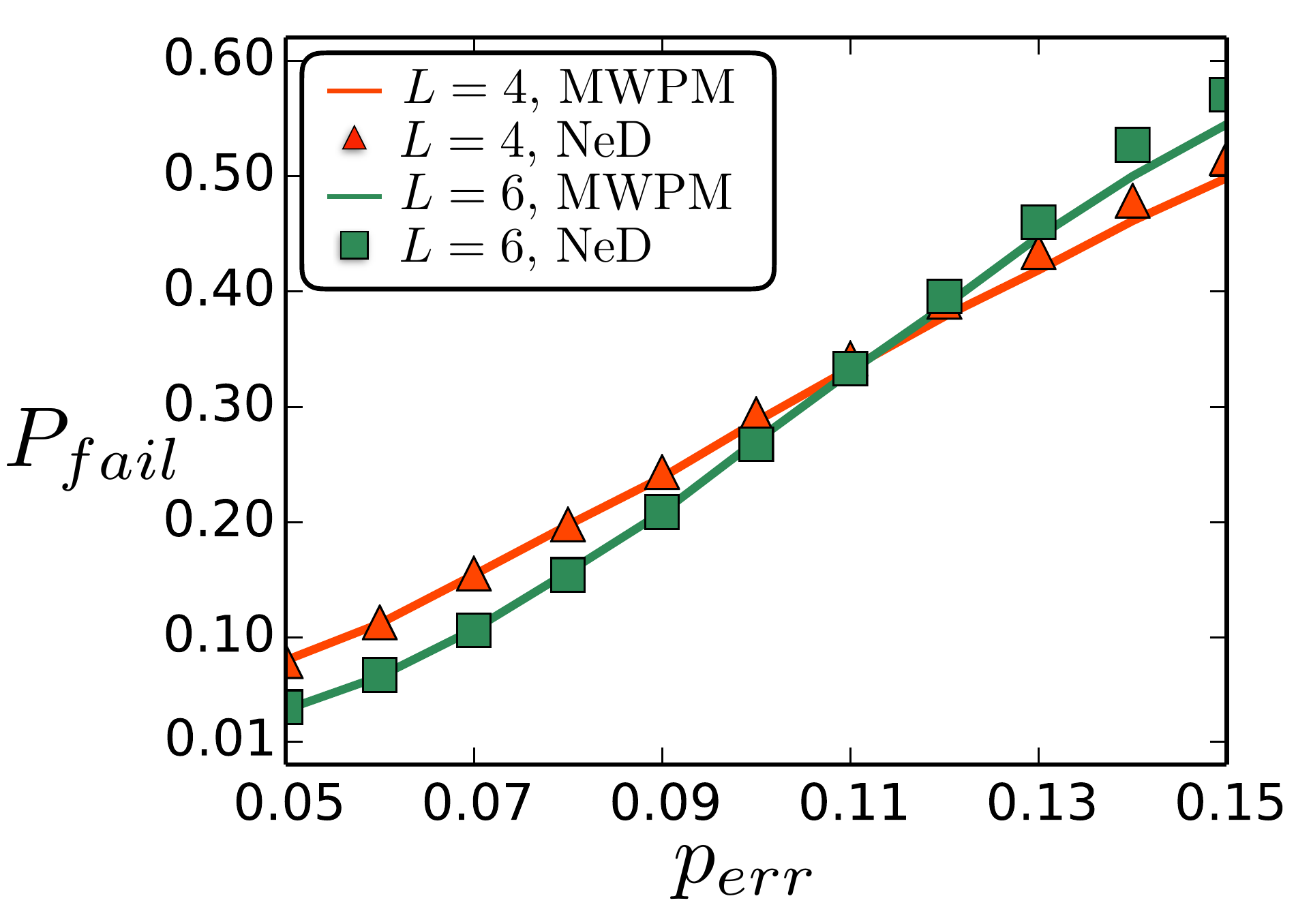}
\caption{Logical failure probability as a function of elementary error probability for MWPM (lines) and the neural decoder (markers) of size $L=4$ (red) and $L=6$ (green).}
\label{accuracy}
\end{figure}
Importantly, error recovery with $\bm{r}^{(k)}$ chosen from the first compatible chain means that the cycle $\bm{e}_k+\bm{r}^{(k)}$ is sampled 
from a distribution that includes all homology classes.
By computing the Wilson loops on the cycles we can measure their homology class.  This allows us to gauge the
accuracy of the decoder in term of the logical failure probability, defined as $P_{fail} = \frac{n_{fail}}{M}$ where $n_{fail}$ is the number of cycles with non-trivial homology.
Because of the fully-connected architecture of the network, and the large complexity of the probability distribution arising from the high degeneracy of error chains given a syndrome, we found that the dataset size required to accurately capture the underlying statistics must be relatively large ($|\mathcal{D}_p|\propto10^5$).
In Fig.~\ref{accuracy} we plot the logical failure probability $P_{fail}$ as a function of the elementary error probability for the neural decoding scheme.

To compare our numerical results we also perform error correction using the recovery scheme given by MWPM~\cite{Kolmogorov02}.
This algorithm creates a graph whose vertices corresponds to the syndrome
and the edges connect each vertex with a weight equal to the Manhattan distance (the number of links connecting the vertices in the original square lattice). MWPM then finds an optimal matching of all the vertices pairwise using the minimum weight, which corresponds to the minimum number of edges in the lattice~\cite{Fowler12}. 
Fig.~\ref{accuracy} displays the comparison between a MWPM decoder (line) and our neural decoder (markers).  As is evident, the neural decoder 
has an almost identical logical failure rate for error probabilities below the threshold ($p_{err} \approx 10.9$~\cite{Dennis02}), yet a significant higher probability above. 
Note that by training the Boltzmann machine on different datasets we have enforced in the neural decoder a dependence on the error probability. This is in contrast to MWPM which is performed without such knowledge.
Another key difference is that the distributions learned by the Boltzmann machine contain the entropic contribution from the high degeneracy of error chains, which is directly encoded into the datasets.
It will be instructive to explore this further, to determine whether the differences in Fig.~\ref{accuracy} come from inefficiencies in the training, the different decoding model of the neural network, or both.  Finite-size scaling on larger $L$ will allow calculation of the threshold defined by the neural decoder.  

\begin{figure}[t]
\centering
\includegraphics[width=70mm]{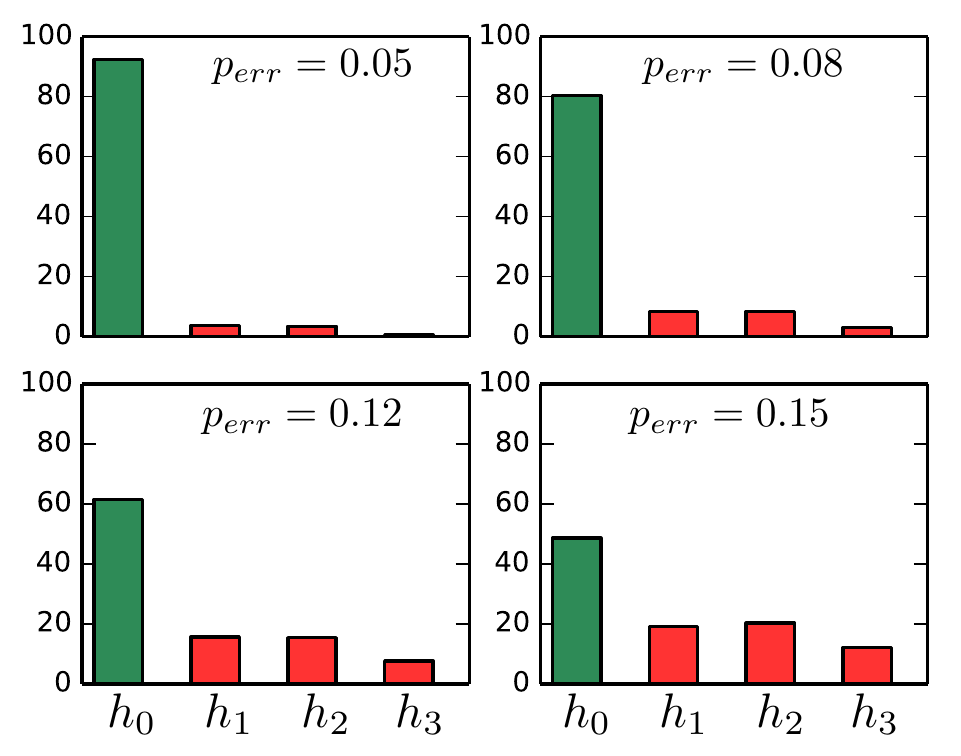}
\caption{Histogram of the homology classes returned by our neural decoder for various  elementary error probabilities $p_{err}$. The green bars represent the trivial homology class $h_0$ corresponding to contractable loops on the torus. The other three classes correspond respectively to the logical operations $\hat{Z}_L^{(1)}$, $\hat{Z}_L^{(2)}$ and $\hat{Z}_L^{(1)}\hat{Z}_L^{(2)}.$
}
\label{histogram}
\end{figure}

In the above algorithm, which amounts to a simple and practical implementation of the neural decoder, our choice to use the first compatible chain for 
error correction means that the resulting logical operation is sampled from a distribution that includes all homology classes.  This is illustrated in 
Fig.~\ref{histogram}, where we plot the histogram of the homology classes for several different elementary error probabilities. 
Accordingly, our neural decoder can easily be modified to perform Maximum Likelihood (ML) optimal decoding.
For a given syndrome, instead of obtaining only one error chain to use in decoding, one could sample many error chains and build up the 
histogram of homology classes with respect to any reference error state.  Then, choosing the recovery chain from the largest histogram bin will implement,
by definition, ML decoding.  Although the computational cost of this procedure will clearly be expensive using the current fully-connected restricted Boltzmann machine, it would be interesting to explore specializations of the neural network architecture in the future to see how its performance may compare to other ML decoding algorithms~\cite{Bravyi14}

{\it Conclusions.} 
We have presented a decoder for topological codes using a simple algorithm implemented with a restricted Boltzmann machine, a common neural network used in many machine learning applications.  
Our neural decoder is easy to program using standard machine learning software libraries and training techniques, and relies on the 
efficient sampling of error chains distributed over all homology classes.
Numerical results show that our decoder has a logical failure probability that is close to MWPM, but not identical,
a consequence of our neural network being trained separately at different elementary error probabilities.  
This leads to the natural question of the relationship between the neural decoder and optimal decoding, 
which could be explored further by a variation of our algorithm that implements maximum likelihood decoding.

In its current implementation, the Boltzmann machine is restricted within a given layer of neurons, but fully-connected between layers.
This means that our decoder does not depend on the specific geometry used to implement the code, nor on the structure of the stabilizer group; it is trained simply using a raw data input vector, with no information on locality or dimension.  In order to scale up our system sizes on the $2D$ toric code
(as required e.g.~to calculate the threshold),
one could relax some of the general fully-connected structure of the network, and specialize it to accommodate the specific details of the code.  This
specialization should be explored in detail, before comparisons of computational efficiency can be made between our neural decoder, MWPM, and other 
decoding schemes.
Note that, even with moderate specialization, the neural decoder as we have presented above can immediately be extended to other choices of error models~\cite{Novais13}, such as the more realistic case of imperfect syndrome measurement~\cite{Wang03}, or transferred to other topological stabilizer codes, such as color codes~\cite{Katzgraber09,Brown16}.

Finally, it would be interesting to explore the improvements in performance obtained by implementing standard tricks in machine learning, such as convolutions, adaptive optimization algorithms, or the stacking of multiple Boltzmann machines into a network with deep structure.
Given the rapid advancement of machine learning technology within the world's information industry, we expect that such tools will be 
the obvious choice for the real-world implementation of decoding schemes on future topologically fault-tolerant qubit hardware.

{\it Acknowledgements.}~The authors thank J. Carrasquilla, D. Gottesman, M. Hastings, C. Herdmann, B. Kulchytskyy, and M. Mariantoni for enlightening discussions.
This research was supported by NSERC, the CRC program, the Ontario Trillium Foundation, the Perimeter Institute for Theoretical Physics,
and the National Science Foundation under Grant No.~NSF PHY-1125915.
  Simulations were performed on resources provided by SHARCNET. Research at Perimeter Institute is supported through Industry Canada and by the Province of Ontario through the Ministry of Research \& Innovation. 

\bibliography{mybib}{}

\section{Supplementary Material}
{\it Training the Boltzmann Machine.} 
We have seen that the training of the neural network consists in finding a set of parameters $\bm{\lambda}$ which minimizes the distance between the dataset distribution and the model distribution $p_{\bm{\lambda}}(\bm{e},\bm{S})$. This probability distribution is obtained from the distribution over the full graph by integrating out the hidden degrees of freedom
\begin{equation}
p_{\bm{\lambda}}(\bm{e},\bm{S})=\sum_{\bm{h}}p_{\bm{\lambda}}(\bm{e},\bm{S},\bm{h})=\frac{1}{Z_{\bm{\lambda}}}\mbox{e}^{-\mathcal{E}_{\bm{\lambda}}(\bm{e},\bm{S})}
\end{equation}
Such marginalization can be carried out exactly, leading to an effective energy
\begin{equation}
\begin{split}
\mathcal{E}_{\bm{\lambda}}(\bm{e},\bm{S})&=-\sum_j b_j e_j-\sum_k d_k S_k+\\&\quad\,-\sum_i\log\left(1+\mbox{e}^{c_i+\sum_kU_{ik}S_k+\sum_j W_{ij}e_j}\right)
\end{split}
\end{equation}
The function to minimize is the average of the KL divergence on the dataset samples, which can be written, up to constant entropy term, as
\begin{equation}
\mathbb{KL}(p_{data}\,|\,p_{\bm{\lambda}})=-\frac{1}{|\mathcal{D}|}\sum_{\{\bm{e},\bm{S}\}} \log p_{\bm{\lambda}}(\bm{e},\bm{S})
\end{equation}
The gradient of the KL divergence thus reduces to the gradient of the log probability
\begin{equation}
\begin{split}
\nabla_{\bm{\lambda}_j}\log p_{\bm{\lambda}}(\bm{e},\bm{S}) &= -\nabla_{\bm{\lambda}_j}\mathcal{E}_{\bm{\lambda}}(\bm{e},\bm{S}) +\\
&\quad\,+ \sum_{\bm{e},\bm{S}}\log p_{\bm{\lambda}}(\bm{e},\bm{S})\nabla_{\bm{\lambda}_j}\mathcal{E}_{\bm{\lambda}}(\bm{e},\bm{S})
\end{split}
\end{equation}
For instance, for the case of the derivative with respect to the weights matrix $\bm{W}$, the gradient of the effective energy is equal to the correlation matrix averaged over the conditional probability of the hidden layer
\begin{equation}
\nabla_{\bm{W}}\mathcal{E}_{\bm{\lambda}}(\bm{e},\bm{S}) = -\sum_{\bm{h}}p_{\bm{\lambda}}(\bm{h}\,|\,\bm{e},\bm{S})\,\bm{e}\,\bm{h}^\top
\label{nablaE_U}
\end{equation}
Therefore the gradient of the KL divergence with respect to $\bm{W}$ can be written as
\begin{equation}
\nabla_{\bm{W}}\mathbb{KL} = -\langle \bm{e}\,\bm{h}^\top \rangle_{p_{\bm{\lambda}}(\bm{h}\,|\,\bm{e},\bm{S})}+\langle \bm{e}\,\bm{h}^\top \rangle_{p_{\bm{\lambda}}(\bm{h},\bm{e},\bm{S})}
\label{nablaKL_W}
\end{equation}
We note now that, due to the restricted nature of the Boltzmann machine (no intra-layer connections), all the conditional probabilities factorize over the nodes of the corresponding layer and can be exactly calculated using Bayes theorem. For instance, the hidden layer conditional distribution factorize as
$p_{\bm{\lambda}}(\bm{h}\,|\,\bm{e},\bm{S})=\prod_{i}p_{\bm{\lambda}}(h_i\,|\,\bm{e},\bm{S})$, where each hidden nodes is activated with probability
\begin{equation}
p_{\bm{\lambda}}(h_i=1\,|\,\bm{e},\bm{S})=\left(1+\mbox{e}^{c_i+\sum_kU_{ik}S_k+\sum_j W_{ij}e_j}\right)^{-1}
\end{equation}
In computing the gradient of the KL divergence, the first average of the correlation matrix in Eq.~\ref{nablaKL_W} is trivial since, given the state of the error layer $\bm{e}\in\mathcal{D}$, we can easily sample the hidden state $\bm{h}$ with the above conditional probability. On the other hand, the second term involves an average over the full probability distribution $p_{\bm{\lambda}}(\bm{h},\bm{e},\bm{S})$, whose partition function is not know and thus inaccessible. To calculate such average correlations we instead run a Markov chain Monte Carlo for $\kappa$ steps
\begin{equation}
\{\bm{e},\bm{S}\}^{(0)}\rightarrow\bm{h}^{(0)}\rightarrow\dots\rightarrow\{\bm{e},\bm{S}\}^{(\kappa)}\rightarrow\bm{h}^{(\kappa)}
\end{equation}
Because $\{\bm{e},\bm{S}\}^{(0)}\in\mathcal{D}$ and thus already belongs to the distribution we are sampling from, there is no need of running a long chain. This algorithm, given the number of steps $\kappa$ of the Markov chain, is called contrastive divergence (CD$_\kappa$)~\cite{Hinton02}. The optimization algorithm used to update the parameters $\bm{\lambda}$  is called stochastic gradient descent. Instead of evaluating the average gradient on the entire dataset, we divide $\mathcal{D}$ into mini-batches $\mathcal{D}^{[b]}$ and update $\bm{\lambda}$ for each mini-batch $b$
\begin{equation}
\bm{\lambda}_j\leftarrow\bm{\lambda}_j-\frac{\eta}{|\mathcal{D}^{[b]}|}\sum_{\{\bm{E},\bm{S}\}\in\mathcal{D}^{[b]}}\nabla_{\bm{\lambda}_j}\mathbb{KL}\,(p_{data}\,||\,p_{\bm{\lambda}})
\end{equation}
where the step $\eta$ of the gradient descent is called learning rate. The initial values of the parameters are drawn from an uniform distribution centered around zero with some width $w$. We also note that a common issue arising in the training of neural networks is the overfitting, i.e. the network reproducing very well the distribution contained in the training dataset but being unable to properly generalize the learned features. To avoid overfitting we employ weight-decay regularization, by adding an extra penalty term to the KL divergence, proportional to the square weights times a coefficient $\mathcal{L}_2$~\cite{Krogh92}. It is now clear that, in addition to the network parameters $\bm{\lambda}$, there are several hyper-parameters dictating the performance of the training. Specifically, the hyper-parameters are the learning rate $\eta$, the size of the mini-batches $b_S$, the width $w$ of the distribution for the initial values of the weights, the order $\kappa$ in the contrastive divergence algorithm , the amplitude $\mathcal{L}_2$ of weight decay regularization and the number of hidden units $n_h$. To find a suitable choice of these external hyper-parameters, we perform a grid search where we train different networks for several combinations of such parameters. We then select the network with higher performance in terms of logical failure probability evaluated over a reference set of error chains.

\end{document}